\documentclass[12pt]{article}
\usepackage{arxiv}
\usepackage[utf8]{inputenc}
\usepackage[T1]{fontenc}
\usepackage[greek,english]{babel}
\usepackage{amsmath}
\usepackage{hyperref}
\usepackage{url}
\usepackage{booktabs}
\usepackage{amsfonts}
\usepackage{natbib}
\usepackage{multirow,enumitem}
\usepackage{array}
\usepackage{algorithm}
\usepackage{authblk}
\usepackage[noend]{algpseudocode}
\usepackage{tikz}
\usetikzlibrary{arrows.meta,positioning,fit,backgrounds,calc}
\usepackage{listings}

\newcounter{algo}

\newcolumntype{L}[1]{>{\raggedright\let\newline\\\arraybackslash\hspace{0pt}}m{#1}}
\newcolumntype{C}[1]{>{\centering\let\newline\\\arraybackslash\hspace{0pt}}m{#1}}
\newcolumntype{R}[1]{>{\raggedleft\let\newline\\\arraybackslash\hspace{0pt}}m{#1}}

\title{\textsc{Quanta}: A Self-Contained Python Library for Hybrid Retrieval over Quantised Embeddings, Lexical Indexes, and Knowledge Graphs}

\author[1,2]{\href{https://orcid.org/0000-0002-3996-3301}{Ioannis E. Livieris}}

\affil[1]{Novelcore, Athens, GR 10436}

\affil[2]{Department of Business Administration \& Organization Administration\\
	 University of Peloponnese, Kalamata,
	 GR 24100 \texttt{livieris@uop.gr}}

\renewcommand{\shorttitle}{\normalsize{\textsc{Quanta}: A Self-Contained Python Library for Hybrid Retrieval over Quantised Embeddings, Lexical Indexes, and Knowledge Graphs}}

\hypersetup{
	pdftitle={\textsc{Quanta}: A Self-Contained Python Library for Hybrid Retrieval over Quantised Embeddings, Lexical Indexes, and Knowledge Graphs},
	pdfsubject={cs.CL, cs.AI},
	pdfauthor={Ioannis E. Livieris},
	pdfkeywords={Hybrid retrieval, rank fusion, knowledge graphs, vector quantisation, retrieval-augmented generation},
}

\begin{document}
    \maketitle

\begin{abstract}
An advanced retrieval-augmented generation pipeline is typically assembled from three or four independently operated systems: an approximate nearest-neighbour index, a full-text search engine, a graph database, and a relational document store. Each contributes its own deployment surface, configuration model, and failure modes, and the integration logic that binds them is written anew in every project.
In this work, we present \textsc{Quanta}, an open-source Python library, which unifies dense
vector search over 4-bit quantised embeddings, BM25 full-text retrieval, and
knowledge-graph traversal behind a single retrieval API. Quanta makes two design
commitments, which distinguish it from existing hybrid retrieval stacks. First,
signals are combined by \emph{weighted reciprocal rank fusion} rather than by
normalising heterogeneous scores onto a shared range, which we argue is
ill-posed because such normalisations are query-dependent. Second, the graph is
a \emph{candidate expander and not a relevance scorer}: traversal widens the
candidate pool, and the newly admitted documents are re-scored by the dense
indexes under an identifier allowlist, so structural adjacency determines what
is considered while content evidence determines how it ranks. Every optional
component---graph, lexical index, embedding cache---has a null implementation,
so a deployment can begin as pure vector search and acquire further signals
through configuration alone. We describe a running deployment---a tool-using
language-model agent over a clinical knowledge graph---and report its observed
behaviour. This is a system description; a controlled retrieval evaluation is
left to future work.
\end{abstract}

\keywords{Hybrid retrieval\and knowledge graphs\and vector quantisation\and retrieval-augmented generation.}

\section{Introduction}
\label{sec:intro}

Retrieval-augmented generation (RAG)~\cite{gao2024ragsurvey,lewis2020rag} has
made retrieval quality a first-order concern for applications previously served
by a single similarity index, and no single signal is sufficient. Dense
bi-encoders~\cite{karpukhin2020dpr} generalise across
vocabulary mismatch but degrade on rare tokens such as product codes, statute
numbers and proper nouns. Lexical scoring with BM25~\cite{robertson2009bm25}
covers that gap, and hybrid dense--lexical retrieval is now standard practice.
Neither signal exploits relationships known to the corpus owner yet absent from
the text: that a court decision cites a statute, or that a diagnosis is
clinically comorbid with another.

Combining all three signals is, in current practice, an integration problem
rather than a retrieval problem. A representative stack pairs
FAISS~\cite{johnson2019faiss} or a managed vector
database~\cite{wang2021milvus} with Elasticsearch for BM25,
Neo4j~\cite{neo4j} for traversal and PostgreSQL for storage; four systems must
be provisioned and kept consistent, and the code fanning a query across them is
rewritten in each project. Orchestration frameworks~\cite{llamaindex}
reduce this boilerplate but do not remove the systems, and treat the graph,
where present, as a separate index with its own query path rather than as a
participant in one ranking.

Quanta delegates vector compression entirely, storing vectors through
\textsc{turbovec}, an implementation of TurboQuant~\cite{turboquant}; it claims no
contribution to quantisation or approximate
indexing~\cite{malkov2020hnsw}. Its fusion rule is likewise
established: reciprocal rank fusion (RRF)~\cite{cormack2009rrf} combines ranked
lists without the score normalisation CombSUM-style
fusion~\cite{fox1994combination} requires, and Bruch \textit{et
	al.}~\cite{bruch2023fusion} show that convex combination surpasses RRF only when
tuned per collection, leaving RRF the stronger default in the untuned regime a
general-purpose library must assume. The principal departure from prior work
concerns the graph: GraphRAG and
related systems~\cite{edge2024graphrag} extract an entity graph using a language
model and apply it to query-focused summarisation, whereas Quanta consumes a
graph authored by the corpus owner, inside ordinary top-$k$ retrieval, and
restricts its influence to candidate selection. The contributions of this paper are threefold:

\begin{itemize}
	\item a single-process retriever fusing dense, lexical and structural evidence
	over one document store, with no external retrieval service
	(Section~\ref{sec:arch});
	\item a two-pass procedure in which graph traversal enlarges the candidate set
	and the admitted documents are re-scored by the dense indexes through an
	allowlist, so that the graph alters what is retrieved without altering the
	ranking function (Sections~\ref{sec:fusion} and~\ref{sec:graph}); and
	\item measurements from a clinical knowledge-graph deployment characterising
	candidate enlargement, fusion behaviour and cost (Section~\ref{sec:case}).
\end{itemize}

The remainder of this paper is organised as follows. Section~\ref{sec:arch}
describes the system architecture, including the quantised vector indexes,
document stores, rank-fusion mechanism and graph-based candidate expansion.
Section~\ref{sec:case} presents a case study on a clinical knowledge-graph
deployment and characterises candidate enlargement, retrieval behaviour and
runtime cost. Finally, Section~\ref{sec:concl} discusses the limitations of the
current design and concludes the paper.

\section{System Architecture}\label{sec:arch}

A deployment comprises a set of named vector indexes, a document store and any
subset of three optional backends: a graph, a lexical index and an embedding
cache. A single \textsc{MultiRetriever} owns them and exposes one asynchronous
\textsc{search()} entry point.

\begin{figure}[!ht]
	\centering
	\includegraphics[width=1.0\textwidth]{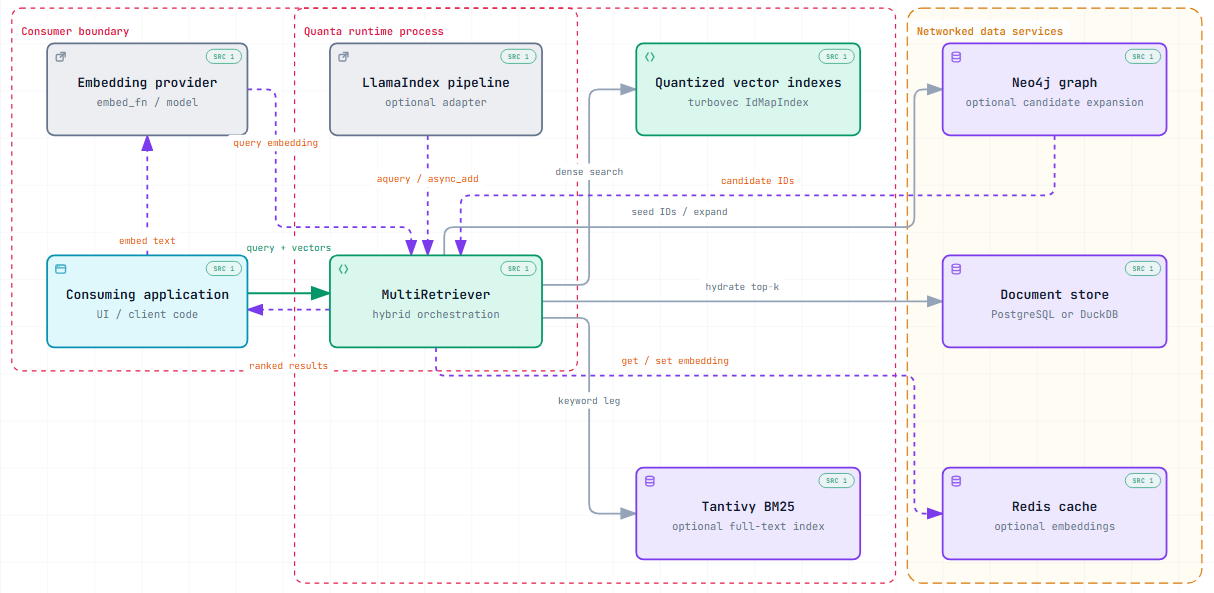}
	\caption{Quanta runtime architecture. The graph returns candidate identifiers
		rather than scores, and the document store is contacted once, to hydrate the
		ranked top-$k$.}
	\label{fig:runtime}
\end{figure}

Fig.~\ref{fig:runtime} shows the runtime decomposition. Quanta performs no
embedding of its own, so the embedding provider remains on the consumer side and
an optional LlamaIndex adapter may drive the same retriever. Components that
must be local---the quantised vector indexes and the Tantivy full-text
index---reside in the library process, whereas the document store, the Neo4j
graph and the Redis cache are networked services. The edges leaving the
retriever are the legs of a query: a dense search per index, a keyword leg, a
seed-and-expand exchange with the graph returning candidate \emph{identifiers}
rather than scores, and one hydration of the ranked top-$k$.

\subsection{Index and Storage Layer}

\textsc{QuantaIndex} wraps a \textsc{turbovec} identifier-mapped index. Application
identifiers are strings whereas the index addresses vectors by 64-bit integers;
Quanta maps between them with a fixed-seed \texttt{xxhash}-64 digest and rejects
insertions whose digest collides with a different existing identifier. At bit
width $b$, an index of $n$ vectors of dimension $d$ occupies $ndb/8$ bytes: for
$10^6$ vectors of dimension 768 at $b=4$ this is 366~MiB against 2.86~GiB for
float32, obtained in-process with no ANN service to operate. Search is exact
over the quantised representation, so query time is linear in corpus size.

Two document-store backends implement one asynchronous interface:
\textsc{PostgresDocStore}, using \textsc{asyncpg} with a GIN index over a
\textsc{JSONB} metadata column, and the embedded
\textsc{DuckDBDocStore}~\cite{raasveldt2019duckdb}. Metadata filtering is
commonly implemented as a post-filter, retrieving the top-$K$ vectors and
discarding those failing the predicate, which under a selective predicate
shrinks the effective $K$ and costs recall invisibly. Quanta resolves the
predicate first, obtains the identifiers satisfying it and passes them to the
index as an allowlist, so the nearest-neighbour scan ranges only over admissible
vectors. The same allowlist enables the re-scoring of
Section~\ref{sec:graph}; both features rest on one primitive.

\subsection{Rank Fusion}
\label{sec:fusion}

The legs emit incommensurable quantities. Cosine similarity is bounded and, for
a competent encoder over a topical corpus, concentrated near the upper end of
its range; BM25 is unbounded and grows with query length and collection
statistics; inverse hop distance is a property of the graph and the seed set
rather than of the query. Quanta therefore fuses on rank. For legs
$\ell \in \mathcal{L}$ with weights $w_\ell$, where $r_\ell(d)$ is the one-based
position of document $d$ in the ranked list of leg $\ell$,

\begin{equation}
	\mathrm{score}(d) = \sum_{\ell \in \mathcal{L}}
	\frac{w_\ell}{k_{\mathrm{rrf}} + r_\ell(d)},
	\label{eq:rrf}
\end{equation}

\noindent where a document absent from a leg contributes nothing from it and
$k_{\mathrm{rrf}}$ (default 60) damps the head of each list. Each dense index
carries $w_{\mathrm{dense}}/|\mathcal{I}|$ over the active index set
$\mathcal{I}$, the lexical leg $w_{\mathrm{bm25}}$ and the graph leg
$w_{\mathrm{graph}}$. Since (\ref{eq:rrf}) is homogeneous in the weights, only
their ratios are meaningful and no renormalisation is required.

\subsection{Graph Expansion as Candidate Enlargement}
\label{sec:graph}

Retrieval proceeds in two passes (Algorithm~\ref{alg:pipeline}). The first ranks
the corpus by content alone and a preliminary application of (\ref{eq:rrf})
selects the top \textsc{graph\_seed\_k} documents as seeds; a bounded
breadth-first traversal then returns every document within \textsc{graph\_hops}
steps of a seed. The admitted identifiers are not injected with a synthetic
score but re-scored: a second search per dense index, constrained through the
allowlist to exactly those identifiers, is merged into that index's ranked list
before the final fusion. An expanded document therefore competes on its own
similarity to the query, on the same footing as one the first pass returned.

Let $\mathcal{C}_1$ denote the candidate set of the first pass and
$\mathcal{C}_2 \supseteq \mathcal{C}_1$ the set after expansion. With
$w_{\mathrm{graph}} = 0$ the graph leg vanishes from (\ref{eq:rrf}), so a
document's fused score depends only on its ranks in the content legs and
traversal effects $\mathcal{C}_1 \mapsto \mathcal{C}_2$ and nothing further. The
graph determines which documents are considered, never how well they rank.

\begin{algorithm}[!ht]
	\caption{Two-pass hybrid retrieval}
	\label{alg:pipeline}
	\begin{algorithmic}[1]
		\Require Query embedding $q$, query text $t$, active indexes $\mathcal{I}$,
		allowlist $A$, fetch size $K$, graph seed size $S$, graph hops $H$
		\Ensure Ranked top-$k$ documents
		
		\Statex
		\State \textbf{Pass 1: Content retrieval}
		\ForAll{$i \in \mathcal{I}$}
		\State $L_i \gets
		\operatorname{rank}(\operatorname{search}(i,q,K,A))$
		\EndFor
		\State $L_{\mathrm{bm25}} \gets
		\operatorname{rank}(\operatorname{search}(\mathrm{BM25},t,K))$
		\State $R_1 \gets
		\operatorname{RRF}(\{L_i\}_{i\in\mathcal{I}},L_{\mathrm{bm25}})$
		\State $Seeds \gets \operatorname{top}_S(R_1)$
		
		\Statex
		\State \textbf{Pass 2: Graph expansion and constrained re-scoring}
		\State $L_{\mathrm{graph}} \gets
		\operatorname{expand}(\mathrm{Graph},Seeds,H)$
		\State $N \gets
		\operatorname{keys}(L_{\mathrm{graph}})
		\setminus \operatorname{covered}(\{L_i\}_{i\in\mathcal{I}})$
		\ForAll{$i \in \mathcal{I}$}
		\State $L'_i \gets
		\operatorname{search}(i,q,\operatorname{allowed\_ids}=N)$
		\State $L_i \gets \operatorname{merge}(L_i,L'_i)$
		\EndFor
		
		\Statex
		\State \textbf{Final fusion}
		\State $R \gets
		\operatorname{RRF}(\{L_i\}_{i\in\mathcal{I}},
		L_{\mathrm{bm25}},L_{\mathrm{graph}})$
		\State \Return $\operatorname{top}_k(R)$
	\end{algorithmic}
\end{algorithm}

Setting $w_{\mathrm{graph}} > 0$ admits the graph as a further leg ranked by hop
distance. This is warranted only when expanded documents possess no vector in
any index and therefore cannot be re-scored, because (\ref{eq:rrf}) compresses
each leg severely. With one dense index, $w_{\mathrm{dense}} = 0.7$,
$k_{\mathrm{rrf}} = 60$ and the frequently quoted setting
$w_{\mathrm{graph}} = 0.3$, a document ranked first by content scores
$0.7/61 = 0.0115$, whereas one ranked fourth by content but first in the graph
leg---an arbitrary one-hop neighbour of a seed---scores
$0.7/64 + 0.3/61 = 0.0159$ and overtakes it. A graph weight within a small
factor of the content weight is therefore not a tie-breaker: it promotes every
one-hop neighbour above the best content match. Quanta defaults to
$w_{\mathrm{graph}} = 0$ and directs tuning towards \textsc{graph\_seed\_k} and
\textsc{graph\_hops}, which widen the pool without perturbing the order.

\section{Case Study: A Clinical Knowledge Graph}
\label{sec:case}

The following characterises observed behaviour on a deployment distributed with
the library. It is not a retrieval evaluation: no relevance judgements and no
baseline ranker are involved.

\begin{figure}[!ht]
	\centering
	\includegraphics[width=1.0\textwidth]{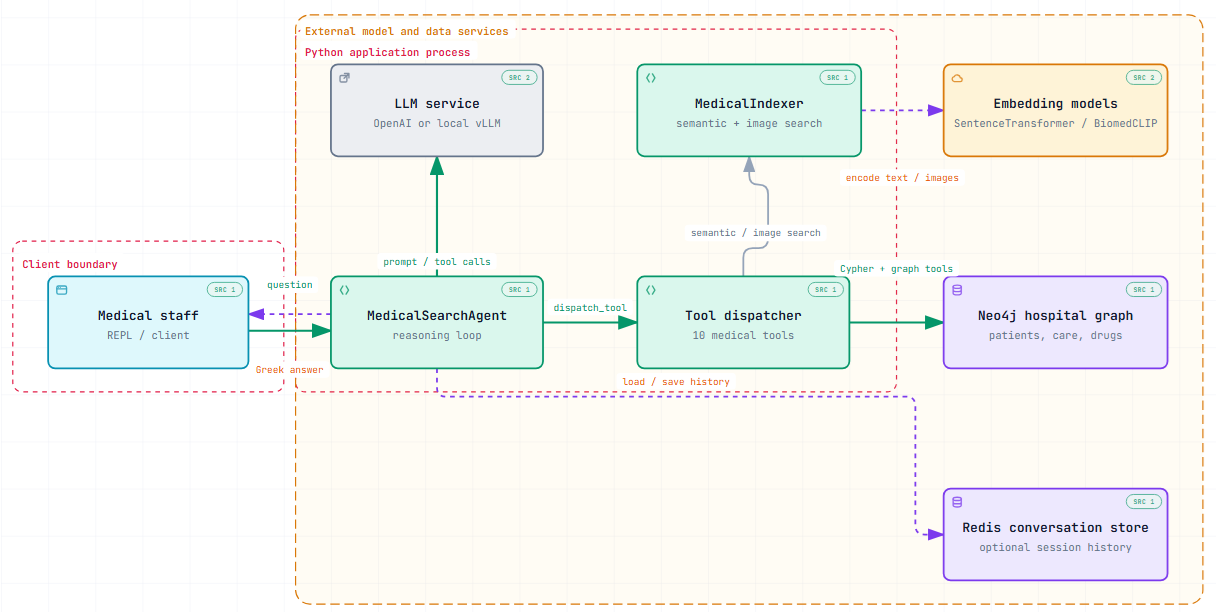}
	\caption{Medical search agent runtime. The dispatcher is the only component that
		reaches data; the quantised indexes reside inside the application process.}
	\label{fig:agent}
\end{figure}

\subsection{Corpus and Ontology}

The corpus is a synthetic hospital record set in Neo4j comprising 100 patients,
30 physicians, 12 diagnoses coded to ICD-10, 12 procedures, 12 medications and 5
hospitals, joined by nine relationship types. The ontology places properties
that vary per relationship on the edges rather than the nodes: a
\textsc{HAS\_DIAGNOSIS} edge carries the severity and chronicity of that diagnosis
for that patient, so the same diagnosis is mild for one patient and severe for
another. Two edge families encode knowledge no document states: nine
\textsc{COMORBID\_WITH} edges between diagnoses, graded by evidence level, and ten
\textsc{INTERACTS\_WITH} edges between medications, graded by severity. Because
these edges are authored rather than extracted, the corpus tests graph expansion
instead of restating what the embeddings already encode.

Five narrative fields are embedded into separate indexes: patient summaries
(100 vectors), physician expertise (30), diagnosis and procedure descriptions
(12 each) and medical images (198), using a 768-dimensional multilingual
sentence encoder for text and a 512-dimensional CLIP encoder for images. All
indexes are quantised to four bits, occupying 107.3~KiB against 858~KiB for
float32. Every narrative is written in Greek.

\subsection{Retrieval Behaviour}

Three dense legs---diagnoses, procedures and patient summaries---were registered
with a Tantivy BM25 leg over the same narratives and a graph leg, using
$w_{\mathrm{dense}} = 0.5$, $w_{\mathrm{bm25}} = 0.3$, $w_{\mathrm{graph}} = 0$
and an embedded DuckDB store. The bundled \textsc{Neo4jGraph} assumes a
homogeneous \textsc{(:Document \{id\})} model, which this graph is not; the
deployment therefore implements the \textsc{GraphBackend} interface directly
against the real schema in roughly forty lines, coalescing per-label key
properties into the identifier used by the indexes. No projection or migration
of the graph was required.

Table~\ref{tab:expansion} reports the candidate pool before and after expansion.
Dense retrieval over three indexes yields 18 identifiers; a two-hop traversal
from three seeds reaches 50 nodes, of which 41--42 are new, and approximately
70\% of those receive a similarity from the constrained re-scoring search. The
pool grows by a factor of roughly 3.3 with no document receiving a score
originating in the graph.

\begin{table}[t]
	\caption{Candidate Enlargement, $k=6$, Three Seeds, Two Hops}
	\label{tab:expansion}
	\centering
	\footnotesize
	\setlength{\tabcolsep}{4pt}
	\begin{tabular}{@{}lrrrrr@{}}
		\toprule
		Query & Pool & BFS & New & Re-scored & Pool$'$ \\
		\midrule
		Insulin resistance, polyuria (el) & 18 & 50 & 41 & 28 & 59 \\
		Chronic kidney disease (en)       & 18 & 50 & 42 & 30 & 60 \\
		Heart failure, dyspnoea (el)      & 18 & 50 & 42 & 28 & 60 \\
		\bottomrule
	\end{tabular}
\end{table}

For the query \emph{heart failure and dyspnoea}, the content legs place
\textsc{I50} (heart failure) first and \textsc{E11} (type 2 diabetes) sixth, the
latter through the BM25 leg alone: no dense index ranked it within its top-$k$,
so it carried no dense contribution. Expansion from the seed \textsc{I50} crosses
an authored \textsc{COMORBID\_WITH} edge and admits \textsc{E11}; re-scoring returns
its similarity, yielding rank 6 in the diagnoses leg and a dense term of
$0.1\overline{6}/66 = 0.0025$. Its fused score rises from $0.0046$ to $0.0071$
and it advances from sixth position to fourth. The graph determined that a
diabetes description merited consideration for a heart-failure query; it did not
determine where that description ranked. Dense and dense-plus-BM25 searches
complete in 7--9~ms on this corpus, and enabling expansion raises query time to
44--115~ms.

\subsection{Agent Deployment}

The same substrate supports a tool-using agent exposing twelve tools: dense
retrieval against a \textsc{QuantaIndex}, lexical resolution over Neo4j full-text
indexes with edit-distance matching, and bounded Cypher traversals.
Fig.~\ref{fig:agent} shows the runtime. A clinician's question enters a
reasoning loop alternating between an OpenAI-compatible model endpoint and a
dispatcher, the only component that reaches data: it issues Cypher and graph
tools against Neo4j and semantic or image searches through the indexer, which
calls the sentence and CLIP encoders. Conversation history is persisted to Redis
when configured. Retrieval is entirely in-process; only the model endpoint, the
graph and the cache cross a network boundary.

Six unseen clinical questions were answered by an open-weight 27B model on a
local endpoint, all without error. Tool execution ranged from 19 to 984~ms
against turn latencies of 19--32~s dominated by generation, below five percent
of the turn throughout. This is the regime in which an embedded, exactly
searched, compressed index is an appropriate trade, and in which the in-process
design removes a network round trip from a loop issuing several tool calls. Two
further properties are exercised: the allowlist serves cohort-restricted
similarity by mechanically the same call that re-scores expanded candidates, and
cross-lingual matching survives quantisation, an English referral question
matching a Greek nephrology narrative at cosine 0.620.

\section{Conclusion}\label{sec:concl}

In this work, we present \textsc{Quanta}, a minimal-dependency architecture for hybrid retrieval that combines semantic, lexical, and structural signals within a lightweight Python library. Its central design principle is to separate candidate generation from relevance estimation: the knowledge graph is used primarily to expand the candidate set through bounded structural traversal, while semantic and lexical evidence remains responsible for ranking the retrieved documents. This separation allows structural relationships to improve the coverage of retrieval without allowing graph proximity to dominate content-based relevance.

A second aspect of the design is the use of quantised vector indexes for memory-efficient dense retrieval. Quanta performs exhaustive search over 4-bit representations, providing exact nearest-neighbour search with respect to the quantised vectors while substantially reducing the memory footprint of the vector store. Combined with identifier-based allowlists, the same retrieval primitive supports both metadata-constrained search and the constrained re-scoring of graph-expanded candidates, without requiring a separate approximate-nearest-neighbour service.

The resulting architecture provides a simple integration path from pure dense retrieval to hybrid dense--lexical--graph retrieval. Optional components are exposed through common interfaces and null implementations, allowing deployments to introduce lexical search, graph expansion, caching, or external document storage without changing the retrieval logic itself. Our clinical deployment demonstrates the operational behaviour of this design, including substantial candidate-set enlargement through graph traversal and the relatively small latency contribution of retrieval compared with downstream language-model generation. These observations characterise the architecture rather than establish retrieval superiority; a controlled evaluation against relevance-labelled benchmarks and alternative hybrid rankers remains an important direction for future work. Quanta is MIT-licensed and available
at \url{https://github.com/ilivieris/quanta}.

\bibliographystyle{plain}
\bibliography{bibliography}

\end{document}